\documentclass[notoc]{JHEP3}
\usepackage{amsmath}
\usepackage{epsfig}
\usepackage{amssymb,amsfonts}

\newcommand{\be}{\begin{equation}}
\newcommand{\ee}{{\end{equation}}}
\newcommand{\ba}{\begin{eqnarray}}
\newcommand{\ea}{{\end{eqnarray}}}

\newcommand{\Real}{\mathbb{R}}

\newcommand{\ie}{{\it i.e.~}}

\newcommand{\RR}{\mathbb{R}}
\newcommand{\ZZ}{\mathbb{Z}}

\newcommand{\half}{\frac{1}{2}}

\def\sqr#1#2{{
\vcenter{\vbox{\hrule height.#2pt
\hbox{\vrule width.#2pt height#1pt \kern#1pt
\vrule width.#2pt}
\hrule height.#2pt}}}}
\boldmath
\title{World-sheet duality for D-branes with travelling waves}
\unboldmath
\author{Constantin P. Bachas$^{1,3}$ 
 and Matthias R. Gaberdiel$^{2,3}$
\\
$\ $ \\
$\ $ \\
$^1$Laboratoire de Physique Th{\'e}orique
de l'Ecole Normale Sup{\'e}rieure\thanks{Unit{\'e} mixte  du
CNRS et de l'Ecole Normale Sup{\'e}rieure,
UMR 8549.} \\
\  24 rue Lhomond, 75231 Paris Cedex 05, France\\
\ Email: \email{bachas@corto.lpt.ens.fr} \\
$\ $ \\
$^2$Department of Mathematics, King's College London\\
\ Strand, London WC2R 2LS, United Kingdom \\
\ Email: \email{mrg@mth.kcl.ac.uk}\\
$\ $ \\
$^3$Theoretische Physik, ETH-H\"onggerberg\\
\  CH-8093 Z\"urich, Switzerland\\
\ Email: \email{bachas@phys.ethz.ch},
 \email{gaberdiel@itp.phys.ethz.ch}
}

\abstract{We study D-branes with plane
waves of arbitrary profiles as examples of
time-dependent backgrounds in string theory. 
We show how to reproduce
the quantum mechanical (one-to-one) 
open-string S-matrix 
starting from the closed-string boundary
state for the  D-branes, thereby establishing
 the channel duality of
this calculation. 
The required Wick rotation to a
 Lorentzian worldsheet singles out
as 'prefered'  time coordinate
  the open-string light-cone time.
}

\preprint{October  2003\\ LPTENS 03/25}

\begin{document}

\section{\Large Introduction}

Time dependence has confronted string theorists with some new
technical and conceptual problems. Recently, these
difficulties have become apparent in attempts to understand the
detailed behaviour of the open-string tachyon decay, see for
example \cite{stro,sen,malda}.
Some of these problems can be traced back
to the fact that perturbative string theory is at present formulated
as an S-matrix theory, while asymptotic states cannot always be
defined in time-dependent geometries.
 A related difficulty is that
in calculating amplitudes one must
 perform certain
analytic continuations involving  the target and/or  the world-sheet
times. These can be tricky and may
 require new physical insights.

In a separate development, various authors have analyzed
time-dependent orbifolds involving
 boosts or null boosts rather than
spatial rotations (for a partial list of
 references see [4--16]).
These backgrounds are toy models for a
cosmological bounce,   and they sometimes  involve a
spacelike or null singularity whose fate
in string theory is an important and still
 open problem. The case
of null singularities looks easier  {\it (a)} because in the absence
of infalling matter they are supersymmetric
 and stable \cite{ffs,lms},
and {\it (b)}  because strong gravitational
 effects need not a priori
invalidate the perturbative treatment
 \cite{horpol}. Furthermore, any
stringy resolution of the problem
must surely involve the twisted
 closed-string states,  which are
`light' near the singularity and
 arbitrarily `heavy'  at later times.

It was suggested in \cite{hull} 
(see also \cite{mw,ccl} for related work) that one may
obtain insight into these
issues  by studying an open-string
 analogue of the null
singularities. The relevant brane
 configurations are so-called null
D-brane scissors --- configurations of
 D-branes that intersect in
a null hypersurface. These arise in type-I
 descendants of the parabolic
orbifold, but they can also be embedded in
 a spacetime that is
uncompactified and flat. The advantages of
 working with open strings
are {\it (a)} that we  can isolate
stringy from strong-gravity effects, by
 considering the disc (as opposed
to higher-order)  diagrams 
and {\it (b)} that we can regularise the long-time behaviour by
making the D-branes parallel in the distant
  past and  future
\cite{hull}. This is analogous to stopping
 the expansion of the 'Universe'
in the parabolic-orbifold geometry. 
In this way one  can focus on the physics of stretched open
strings -- the analogue of  twisted closed strings -- near the
singularity, without worrying about strong-gravity effects and
the (non-)existence of asymptotic twisted
states.\footnote{For a discussion  of this latter point
 see  also references  \cite{Nekrasov,Berk}.} 

   The regularised null D-brane scissors are examples of
supersymmetric branes carrying plane-fronted waves.
These,  and various dual configurations,  have been discussed
widely in the literature 
(a partial list of references is [20--26]). 
Recently it was shown in \cite{bachas} that
 for any temporal profiles
of the D-brane  waves, 
the classical and quantum dynamics of an open string 
can be solved exactly. The problem is simple because
 the coupling of the open string to
such plane  waves  only arises through the
  end-points, and  is essentially
linear. This is to be 
contrasted with the case of quadrupolar 
D-brane waves \cite{db},
or of D-branes [29--35]
  in  gravitational pp-waves
\cite{ak,met,prt},  where the world-sheet
 theory is no longer massless,
and in general not even free. 

The analysis of ref. \cite{bachas} was performed
 in open-string 
light-cone gauge. Subsequently, 
a covariant boundary state for a  D-brane with 
travelling waves  has been proposed by the authors of 
\cite{htt} (see also \cite{blum,tak,Call}). 
While there are good reasons to believe that this
boundary state  does indeed describe the relevant
configuration, a direct comparison to the results
 of \cite{bachas} has not
been made. Our aim in this paper is to establish explicitly the
connection, thereby verifying  open/closed-string
 duality in this
specific context.

   Checking channel duality in these
 'partially-chartered waters'
is interesting for many different reasons.
 First, the boundary-state
approach is covariant and can be extended to
 more general amplitudes,
including several  strings and/or  D-branes. The
study of such amplitudes, that we defer to
 future work, could shed
light on the fate of null singularities 
as discussed  in ref. \cite{hull}. 
Secondly, some of the  techniques used in our derivation
(how to deal, for instance,  with normal ordering 
and the Wick rotation)
may  prove useful in other time-dependent backgrounds, 
like  the rolling  tachyons mentioned above. 
 Finally, it is interesting to see how the 
couplings to the bulk closed strings, encoded in the 
conformal boundary state,
capture the non-trivial time history of the D-brane.

\medskip

The paper is organised as follows. In section~2 we review
 the open-string 
calculation of \cite{bachas}, and phrase the resulting 
one-to-one S-matrix
in a way that will prove useful for the comparison with
 the closed-string
calculation. Section~3 reviews ref.  \cite{htt}, giving a 
slightly different derivation of the D-brane boundary  state 
in an oscillator basis.
  Section~4 contains our main result~: we
show how to extract the open-string S-matrix from the 
boundary state, and find  complete agreement
with the original result of \cite{bachas}.
 One subtle point concerns
the Wick rotations~: while the ambient
 spacetime stays throughout
the calculation Lorentzian, the need to
 Wick rotate the
worldsheet   singles  out, as we will see,  
the  open-string light-cone gauge time. Finally 
 we comment
in section~5 on  generalisations of our
 result to other amplitudes.
There is one
appendix on  the precise relation between the two-point
 functions on the
half-cylinder and the infinite strip.


\section{\Large The one-to-one open string S-matrix}

An open string crossing the plane-fronted wave will in
general emerge in a final state containing an arbitrary number $n_o$
of open,  as well as an arbitrary
 number $n_c$ of closed
strings (with $n_o+n_c\geq 1$). The
 corresponding process is weighted
by a power  ${n_o+2n_c-1}$ of 
  the string coupling
constant. Thus, provided  these amplitudes
 do not diverge,  the leading
semiclassical effect is
  the transition of a single open
string from an initial to a final quantum state. The corresponding
S-matrix was calculated in 
 open-string light-cone gauge in
ref.~\cite{bachas}. We here review this  calculation,
both to make the paper self-contained and in order
to express the final result in a convenient form.

The  background consists of a planar  static Dp-brane 
living  in Minkowski
space time and  carrying a plane
 electromagnetic wave $A^i(x^+)$. Here
$x^+,x^-, x^{i=2,\ldots ,  p}$ are 
light-cone coordinates for the
Dp-brane in static gauge,
 and we assume that the wave profile dies out in the past and
future sufficiently fast,
 (what this means will become clear shortly)
\begin{equation}
A^i(x^+) \to 0 \ \ \ {\rm as}\ \ x^+\to \pm\infty\; .
\end{equation}
A T-duality transformation maps this background to an
undulating D-brane with planar  wavefronts. Our analysis
can therefore be carried over to
 this case with only minor (essentially
semantic) modifications.

We consider an open string in this background, and treat
the wave as an interaction term.
The interaction Hamiltonian in the
light-cone gauge reads \footnote{In the NSR formulation of the
superstring the fermionic part of the interaction Hamiltonian is
proportional to $\psi^+$, which vanishes  in light-cone gauge.
Thus the one-to-one open string S-matrix is the same for both
the bosonic and the fermionic strings.
We thank N. Couchoud for this observation.}
\begin{equation}\label{hami}
{\cal H}_I(\rho) = A^i(p^+\rho)
\left( \partial_\rho X^i(0,\rho) -
\partial_\rho X^i(\pi,\rho)\right) \, ,
\end{equation}
where the open string is parametrised by
$\varphi\in [0,\pi]$, the light-cone
time is $X^+= p^+\rho$,
summation over the index $i$ is implicit,
and the mode expansion of a Neumann
coordinate is given by the
standard expression
\begin{equation}\label{openmodes}
X^i(\varphi,\rho) = x^i + p^i\rho +
 \sum_{n\not= 0} \frac{i}{n} \,
a_n^i \, e^{-in\rho}\, {\rm cos}(n\varphi)\ .
\end{equation}
We use units in which  $2\alpha^\prime =1$. 
The two terms in the Hamiltonian (\ref{hami})
 correspond to the
two string endpoints which carry equal and opposite charge.
The modes obey the canonical commutation relations
\begin{equation}
[a^i_n,a^j_m] = n\delta^{ij}\delta_{n,-m}\; , \qquad\qquad
[x^{\mu}, p^{\nu}] = i \eta^{\mu\nu}\; .
\end{equation}
Our metric is $\eta^{\mu\nu}= {\rm diag}(- + \cdots +)$.
In the case of an undulating
 D-brane, one must replace $A^i$ by
a transverse D-brane
coordinate $Y^i/2\pi\alpha^\prime$,
 the $\rho$-derivatives
in (\ref{hami})  by
$\phi$-derivatives, and the Neumann
 by a Dirichlet mode expansion.

In the interaction representation
the quantum-mechanical S-matrix reads
\begin{equation}
\label{sop}
S = {\cal T}\, \exp\left(i\int_{-\infty}^{\infty}{d\rho}\;
{\cal H}_I(\rho)\right) \; ,
\end{equation}
where ${\cal T}$ stands for time ordering with respect to the
light-cone time $\rho$.  We can replace this by normal ordering
at the expense of introducing a real and an imaginary phase,
\begin{equation}\label{smat}
S = e^{i\delta_1}e^{-\delta_2}
: {\rm exp}\left( i\int_{-\infty}^{\infty}{d\rho}\;
{\cal H}_I(\rho)\right):\ \ .
\end{equation}
Because the interaction is linear, the problem
 reduces to that of
free fields interacting with an external 
time-dependent source
(see {\it e.g.} \cite{IZ}). The result reads
\begin{eqnarray}\label{pha}
\delta_1+i\delta_2 = i
\int_{-\infty}^{\infty}d\rho
\int_{-\infty}^{\infty}
{d\rho^\prime}&&  A^i(p^+\rho)
 A^i(p^+\rho^\prime)\;\times\nonumber
\\ &&
\times\;\frac{\partial}{\partial\rho}
\frac{\partial}{\partial\rho^\prime}\;
\left[ D_F^{\rm str}(\rho-\rho^\prime, 0) -
D_F^{\rm str}(\rho-\rho^\prime, \pi)\right]\ ,
\end{eqnarray}
where
\begin{equation}
D_F^{\rm str}(\rho-\rho^\prime, \phi_0) =
\left\langle {\cal T} \left(X(\rho,0)\,
 X(\rho^\prime,\phi_0)\, \right)
\right\rangle
\end{equation}
is the Feynman propagator for two points
on the same ($\phi_0=0$) or on  opposite ($\phi_0=\pi$)
boundaries of the strip. Evaluating the
 propagator explicitly we find
(up to an irrelevant constant)~:
\begin{eqnarray}\label{stripF}
D_F^{\rm str}(\rho-\rho^\prime, \phi_0)&=&\ 
 \theta(\rho-\rho^\prime)
\sum_{n>0} \frac{1}{n} \, e^{i\phi_0 n} \,
e^{in(\rho^\prime-\rho)(1-i\epsilon)} + (\rho\leftrightarrow
\rho^\prime)
\nonumber\\
&=& -\theta(\rho-\rho^\prime)
\; {\rm log}(1- e^{i\phi_0}
 e^{i(\rho^\prime-\rho)(1-i\epsilon)}) +
(\rho\leftrightarrow
\rho^\prime)\ ,
\end{eqnarray}
where
$\theta(y)$ is the Heaviside
step function,  
and  the $i\epsilon$ rotation of the time
axis renders  the infinite sums
absolutely convergent.

  To separate $\delta_1$ from $\delta_2$ let us introduce
the Fourier components of the  wave profile
(our conventions are as in  ref.~\cite{bachas}):
\begin{equation}\label{fourierdef}
{A}^i(x) \equiv \int_{-\infty}^{\infty} \frac{dk}{k}
{\tilde A}^i(k)\; e^{ikx}\ ,
\end{equation}
and define the new variable $\lambda=\rho-\rho^\prime$.
 Inserting
(\ref{fourierdef}) in eq.~(\ref{pha})
 and performing explicitly the
integrations over $\rho+\rho^\prime$
 and one momentum,
leads to the  expression
\begin{equation}\label{expr}
\delta_1+i\delta_2 = -2 \pi i { p^+}\int_{-\infty}^{\infty}
\hspace*{-0.1cm} dk\,
\int_{-\infty}^{\infty} \hspace*{-0.1cm} d\lambda\,\,
{\tilde A}^i(k) \, {\tilde A}^i(-k)\;  e^{ik\lambda p^+} \,
[ D_F^{\rm str}(\lambda, 0)  - D_F^{\rm str}(\lambda, \pi)]\ .
\end{equation}
We have used, in deriving this expression, the fact that
$D_F^{\rm str}$  vanishes in the infinite 
past and future, which makes it possible
to drop boundary contributions when
integrating by parts. Inserting
the sum representation of
the propagator, performing explicitly
 the $\lambda$-integral, and
changing variables from $k$ to $\alpha \equiv n-kp^+$ gives
\begin{equation}\label{combined}
\delta_1 +i\delta_2 =  -\sum_{n>0 \; {\rm odd}}
\frac{8 \pi}{n}\;
\int_{-\infty}^{\infty} \frac{d\alpha}{\alpha-i\epsilon}\;
{\tilde A}^i(\frac{\alpha-n}{p^+})\;
{\tilde A}^i(\frac{n-\alpha}{p^+})\ .
\end{equation}
This is the momentum-space representation of the phase shift.
We can finally use the well-known identity of distributions
\begin{equation}
\frac{1}{\alpha - i\epsilon} = {\cal P}\frac{1}{\alpha} + i\pi
\delta(\alpha)\ ,
\end{equation}
where ${\cal P}$ stands for the principal part, to
get\footnote{The reality condition for the
 Fourier components is
${\tilde A}(-k) = - {\tilde A}(k)^*$.}
\begin{equation}\label{real}
\delta_1 =  -\sum_{n>0 \; {\rm odd}}
\frac{8 \pi}{n}\;{\cal P}
\int_{-\infty}^{\infty} \frac{d\alpha}{\alpha}\;
{\tilde A}^i(\frac{\alpha-n}{p^+})\;
{\tilde A}^i(\frac{n-\alpha}{p^+})\ ,
\end{equation}
and
\begin{equation}\label{imaginary}
\delta_2 =  \sum_{n>0\;{\rm odd}} \frac{8 \pi^2}{n }\;
\Bigl\vert {\tilde A}^i(\frac{n}{p^+})\Bigr\vert^2\ .
\end{equation}
Notice that the imaginary part is positive-definite. 
The above result 
agrees after a T-duality transformation  with eq.~(5.17) of
ref.~\cite{bachas}.

As pointed out in this reference, the probability
of exciting the string from its ground state,
$P_{exc} = 1 - e^{-2\delta_2}$, vanishes in the $p^+\to 0$  limit. 
This is the adiabatic limit in which
  the string surfs smoothly on the
incident pulse and the single-string
 S-matrix approaches 
the identity operator.
More generally, the phase $\delta_2$ is finite provided that the
Fourier transform $\tilde A^i$ can 
 be defined  and  the sum in
(\ref{combined})  converges. It can be checked  that
these conditions are satisfied whenever
 the  wave carries finite total
energy \cite{bachas}.


\section{\Large The boundary state}

The boundary state for a  Dp-brane with
 a travelling wave has been
analyzed recently by Hikida {\it et al}
 \cite{htt}. In this section
we review the results of these authors, both for the sake of
completeness and in preparation  for the
 closed-channel calculation of
the S-matrix in  section 4. Although the final expression for the
boundary state is the same, our derivation  differs somewhat from
that of ref. \cite{htt}.

The starting point is the following 
expression for the boundary state:
\begin{equation}\label{ansatz}
|\!| {\cal B} \rangle\!\rangle = {\rm P} \exp\left(
- i \int_0^{\pi} d\sigma A^i(X^+) \, \partial_\sigma X^i \right) \,
|\!| {\rm Dp} \rangle\!\rangle \,,
\end{equation}
where $|\!| {\rm Dp} \rangle\!\rangle$ is the  boundary
state for a planar  static Dp-brane, and
the Wilson-loop operator  accounts   for
the background gauge field.
The $X^\mu$'s  in eq. (\ref{ansatz}) are
the closed-string coordinate operators evaluated at
fixed closed-string time $\tau = 0$. Their  mode
expansions for arbitrary $\tau$  take  the usual  form
\begin{equation}
X^\mu(\sigma,\tau) = x_c^\mu + p_c^\mu \tau
+ \frac{i}{2} \sum_{n\ne 0} \frac{1}{n} e^{-2in\tau} \left(
\alpha^\mu_n e^{-2in \sigma} +
\tilde\alpha^\mu_n e^{2in \sigma} \right) \,.
\end{equation}
Here  $\sigma$ runs from
$0$ to $\pi$, and we have added a
 subscript $c$ to the  zero modes to
distinguish them from the ones in the open channel. The
canonical commutation relations are
\begin{equation}\label{modecom}
[\alpha^\mu_n,\alpha^\nu_m] =
 [\tilde\alpha^\mu_n, \tilde\alpha^\nu_m]
= n \eta^{\mu\nu} \delta_{n,-m} \,, \qquad \qquad
[x_c^\mu, p_c^\nu] = i\eta^{\mu\nu}\ .
\end{equation}
It is straightforward to check that these commutation
relations imply   the appropriate
boundary conditions 
\begin{equation}\label{gluing}
\left(\partial_\tau X^\mu - \pi\, F^{\mu\nu}(X^+) \,
\partial_\sigma X_\nu\right)\Bigr\vert_{\tau=0}
|\!| {\cal B}  \rangle\!\rangle = 0\,  , 
\end{equation}
where $F^{\mu\nu}$ is the background gauge-field strength.

  Closer inspection of expression (\ref{ansatz}) reveals that
the path ordering  is trivial
because coordinate fields  and their $\sigma$-derivatives
all commute  at equal world-sheet time.
The non-trivial step taken in \cite{htt} is the
elimination of the positive-frequency (annihilation)
modes of the coordinate fields.
To this end we first write
\begin{equation}
X^\mu(\sigma,0) \equiv  X^\mu_{>}+   X^\mu_{<} + x_c^\mu \,,
\end{equation}
where
\begin{equation}
X^\mu_{>} =
\frac{i}{2} \sum_{n>0} \frac{1}{n}
\left(\alpha^\mu_n e^{-2in\sigma} + \tilde\alpha^\mu_n
e^{2in\sigma}\right)
\end{equation}
with a  similar expression  for $X^\mu_{<}$\ .
 The standard Neumann
boundary conditions read~:
\begin{equation} \label{comb}
(X^\mu_{>} - X^\mu_{<})\; |\!|
 {\rm Dp} \rangle\!\rangle = 0\ \ \
{\rm for}\   {\rm  all}\  \sigma \in (0,\pi]\  .
\end{equation}
The strategy is to factorise the
 Wilson-loop operator in two terms,
one on the left  involving  only  the
 negative-frequency modes, and
one on the right that depends on the
combinations ($X^\mu_{>} - X^\mu_{<}$)
which vanish  when acting on  the state
$ |\!| {\rm Dp} \rangle\!\rangle $. To this end we rewrite 
\begin{eqnarray}\label{ewl}
-i \int_0^\pi d\sigma\; A^i(X^+)\; \partial_\sigma X^i
&=&  -2i \int_0^\pi d\sigma\; A^i(X^+)\;
\partial_\sigma X^i_{<}\; + \nonumber\\
&&+ i \int_0^\pi d\sigma\; A^i(X^+)\; (\partial_\sigma X^i_{<} -
\partial_\sigma X^i_{>}) \,.
\end{eqnarray}
To separate the two exponentials
we use the Baker-Campbell-Hausdorff (BCH) formula
\begin{equation}\label{BCH}
e^{B+C} = e^{B}\, e^{\half [C,B]} \, e^C\ ,
\end{equation}
with 
$B$ and $C$ the first and second
 terms on the right-hand-side
of (\ref{ewl}), respectively.
 The BCH formula is valid whenever both
$B$ and  $C$ commute with their mutual commutator $[C,B]$. 
This is indeed true in our case  
because the transverse coordinates $X^i$ 
enter only linearly in (\ref{ewl}),
while the modes of the
light-cone coordinate $X^+$ commute with everything 
in this expression.
From the BCH formula and the Neumann conditions
(\ref{comb}) we thus obtain
\begin{eqnarray}\label{fin}
|\!| {\cal B} \rangle\!\rangle &=&
\exp\left( -2i \int_0^\pi d\sigma\; A^i(X^+)\;
\partial_\sigma X^i_{<}\right)
\times \\
&&\times \exp\left(- \int_0^\pi d\sigma\; A^i(X^+)\;
\int_0^\pi d\sigma^\prime \; A^i(X^+)\;
\partial_{\sigma}\partial_{\sigma^\prime}
D_F^{\rm cyl}(\sigma,\sigma^\prime)\right)\;
 |\!| {\rm Dp} \rangle\!\rangle
\,, \nonumber
\end{eqnarray}
where
\begin{equation}
D_F^{\rm cyl}(\sigma , \sigma^\prime)\;  =\;
\langle X(\sigma, 0)\, X(\sigma^\prime, 0)\rangle\;
=\; \left[\; X_{>}(\sigma), \;
X_{<}(\sigma^\prime)\; \right]
+ \langle x_c\, x_c \rangle
\end{equation}
is the Feynman propagator on the cylinder evaluated at fixed
closed-string time $\tau=0$. 
To keep the notation light we have suppressed 
in  eqs.~(\ref{ewl}) and (\ref{fin}) the  arguments,
$\sigma$ and $\sigma^\prime$,  
of the coordinate fields. We trust that this will
not  confuse the alert reader.

The two-point function on the cylinder
 can be calculated easily with
the result~:
\begin{eqnarray}\label{comt}
\left[\; X_{>}(\sigma), \;
X_{<}(\sigma^\prime)\; \right] &=&
\sum_{n>0} \frac{1}{4n} \left(e^{2in(\sigma-\sigma')}
+ e^{-2in(\sigma-\sigma')}\right) \nonumber\\
&=&  -\frac{1}{4}\; {\rm  log}\;
\left( 4 \; {\rm sin}^2(\sigma-\sigma^\prime)\right) \ .
\end{eqnarray}
Inserting the infinite-sum representation of $D_F^{\rm cyl}$ in
eq.~(\ref{fin}),  and defining the moments  of the gauge field
\begin{equation}\label{Aexpansion}
A^i(X^+(\sigma,0)) \equiv  \sum_{n\in\ZZ} e^{-2in\sigma} A^i_n \,,
\end{equation}
we arrive at the following expression for the boundary state~:
\begin{equation}\label{fina}
|\!| {\cal B} \rangle\!\rangle =
\exp\left(- 2 \pi^2 \sum_{n>0}n  A^i_{-n} A^i_n \right)
\exp\left(2\pi i \sum_{n>0} (  A^i_{-n} \tilde\alpha^i_{-n}
-  A^i_{n} \alpha^i_{-n})  \right) \,
|\!| {\rm Dp} \rangle\!\rangle \,.
\end{equation}
This agrees,  for $\alpha'=1/2$,  
with the expression  (2.17) of ref.~\cite{htt}.
Note that the definition of the
$A^i_n$  is free of operator-ordering  ambiguities because
the modes of $X^+$ all
commute, and can thus be treated 
in the above  formulae as   c-numbers. 
 
  The positive-frequency modes of $X^+$
 can be actually eliminated
from  (\ref{fina}) with the help of  eq.~(\ref{comb}).
  One immediate
consequence is that the disk partition
 function does not depend on
the wave profile~:\ 
\begin{equation}
Z_{\rm disk}= \langle 0 |\!| {\cal B} \rangle\!\rangle =
\langle 0 |\!| {\rm Dp} \rangle\!\rangle \,,
\end{equation}
where $| 0 \rangle$ is the ground state of the closed string.
It has been argued in \cite{Andr,Witt,Shata}
that $Z_{\rm disk}$ is (at least formally)  proportional
to the on-shell action of
 open-string-field theory. The above result
is therefore  consistent with the fact that the 
 plane-wave background 
solves the classical equations for all profiles.

\section{\Large S-matrix from the boundary state}

  The  boundary state (\ref{ansatz})
describes how undulating  D-branes (or their T-duals)
couple  to the  closed-string fields in the bulk. 
We will now derive from this starting point
the one-to-one open-string S-matrix of section 2. 
The derivation involves, as we will see, a  Wick rotation to
a preferred Lorentzian world-sheet. Our result paves, furthermore,
the way for the  covariant calculation of induced 
open- or closed-string emission processes, on which we will 
briefly comment in the end.

  Let us focus on the forward scattering of
 an open string  prepared
initially in its (tachyonic) ground state,
  and which traverses the
pulse without being excited. Since world-sheet
 fermions do not enter in
the interaction Hamiltonian,
 the calculation is also valid for  
the superstring that is in a state with no bosonic excitations. 
The corresponding amplitude reads~:
\begin{equation}\label{scattering}
S_{0\to 0} = {\cal N}
\; \langle 0| : e^{- i\; p_\mu  X^\mu(0,0)}: \;
: e^{i\; p_\mu  X^\mu(\pi/2 ,0)} : \;
|\!| {\cal B} \rangle\!\rangle  \,,
\end{equation}
where $-p_\mu p^\mu = -1/\alpha^\prime=2$ (we use units with
$2\alpha'=1$) is the mass squared of the open-string tachyon,
the normalisation ${\cal N}$ is independent of  the wave profile, 
and we used the SL(2,$\Real$)
invariance of the disk to fix the vertex-operator insertions  at
antipodal points of the boundary, $\sigma = 0$ and $\sigma =\pi/2$
[recall that  $\sigma$ runs from $0$ to $\pi$].
We want to show that expression (\ref{scattering}) leads  to the
result  of section 2, {\it i.e.}
$S_{0\to 0}= e^{i\delta_1-\delta_2}$ with the phases given
by  eq.~(\ref{pha}). Once we understand how this works, we can,
at least  in principle,
calculate any other element of the quantum mechanical S-matrix
(\ref{smat}) in a similar way.

  The first point that needs to
 be understood is the normal ordering
prescription of the vertex operators in
(\ref{scattering}). If the $X^\mu$ were expanded out in
the open channel, 
we would have to  place their
negative-frequency parts to the left of the
positive-frequency ones. In the closed channel,
 on the other hand, we
must  place $X^\mu_{>}$ to the left of the combination that
annihilates the Dp-brane boundary state~:
\begin{equation}\label{normalord}
:e^{i\; p_\mu  X^\mu}:\  \equiv\
e^{2i\; p_\mu  X^\mu_{<}}\;\; e^{i\; p_\mu  x_c^\mu}\;\;
e^{i\; p_\mu ( X^\mu_{>}- X^\mu_{<})}\,\;.
\end{equation}
This guarantees, as we shall see, the absence of divergences
from self-contractions.\footnote{Alternatively, one
can arrive at  the above
prescription by studying closed-string vertex operators 
as they approach
the boundary of the  disk. For calculations of closed-string
correlation functions in the  boundary-state approach see, 
for example, refs.  
\cite{schulze,watts,kawai}.}
Notice that the  normal ordering
refers to a flat static Dp-brane background. For a pulse that
dies out in the infinite  future and past,
 it is in this background
that one must define the asymptotic states.

  To fix the normalisation ${\cal N}$
we  first calculate (\ref{scattering})
with the pulse switched off. Using the BCH
formula (\ref{BCH}) and the commutator (\ref{comt}) gives~:
\begin{equation}
\langle 0| : e^{-i\; p_\mu  X^\mu(0,0)}: \;
: e^{i\; p_\mu  X^\mu(\sigma ,0)} : \;
 |\!| {\rm Dp} \rangle\!\rangle
=  ( 4\; {\rm sin}^2\sigma)^{- p^2/2 }\;
\langle 0|\!| {\rm Dp} \rangle\!\rangle\, .
\end{equation}
Setting $\sigma=\pi/2$,  and using
 the mass-shell condition $p^2= 2$
we find~: 
\begin{equation}\label{norml}
S_{0\to 0} = \frac{{\cal N}}{4}\; \langle 0|\!| {\rm Dp}
 \rangle\!\rangle\ \  \ \  {\rm (no\ \  pulse)}.
\end{equation}
Limiting ourselves to a sector of given $p^i$
and  $p^+$ (which are conserved since we only consider
strings that are  neutral with respect to $A^i$),   
we may  normalise the S-matrix  so that  $S_{0\to 0} = 1$
when there is no  pulse.
This condition fixes  ${\cal N}$ via  eq. (\ref{norml}).

  Let us  now return to the general expression (\ref{scattering}).
After pushing the negative-mode
 piece of the second vertex operator
to the left of the first vertex
 operator, so as to hit the bra vacuum,
we obtain
\begin{equation}\label{scatter}
S_{0\to 0} = \frac{{\cal N}}{4}
\; \langle 0|  e^{-i\; p_\mu ( X^\mu_> -
 X^\mu_<) }\vert_{\sigma=0} \;
e^{i\; p_\mu  (X^\mu_>- X^\mu_<)}
 \vert_{\sigma={\frac{\pi}{2}}}    \;
|\!| {\cal B} \rangle\!\rangle  \,.
\end{equation}
To further simplify the calculation we will  assume that
$p^i=0$, {\it i.e.} that the
string hits the plane-fronted electromagnetic
wave head-on. This is T-dual 
to the case  of an open string traversing a geometric brane wave, 
as in  \cite{bachas}. The more general situation of an
arbitrary  angle of incidence on the wave, 
can be  analyzed similarly
but would render our expressions  more lengthy. Using
the fact that
\begin{equation}
[\; X^-_{>}(\sigma)- X^-_{<}(\sigma)\; ,\;
X^+_{>}(\sigma^\prime) - X^+_{<}(\sigma^\prime)\;] = 0\ ,
\end{equation}
we can now push the $X^+$ pieces 
 of the two  exponentials in (\ref{scatter}) to the right, 
past the Wilson-loop operator which depends only on  $X^i$ and 
$X^+$. Since $X^+_{>} - X^+_{<}$   annihilates
the Neumann boundary state we are left with~:
\begin{equation}\label{scatter1}
S_{0\to 0} = \frac{{\cal N}}{4}
\; \langle 0|  e^{i\; p^+ ( X^-_> - X^-_<) }\vert_{\sigma=0} \;
e^{-i\; p^+  (X^-_>- X^-_<)} \vert_{\sigma={\frac{\pi}{2}}}    \;
|\!| {\cal B} \rangle\!\rangle  \,.
\end{equation}

  To proceed further we will use the general identity (of which the
BCH formula is a special case)
\begin{equation}\label{bch1}
e^C g(B) e^{-C} = g(B+[C,B]) \,.
\end{equation}
This holds  for an arbitrary  function $g(B)$, 
provided that $[C,B]$ commutes with both $B$ and with $C$.
Letting $B = X^+(\sigma)$ and
$C= i\; p^+  (X^-_>- X^-_<)\vert_{\pi/2}$, and using once again the
commutator (\ref{comt})
we find~:
\begin{eqnarray}\label{rrs}
e^{-i\; p^+  (X^-_>- X^-_<)} \vert_{\sigma={\frac{\pi}{2}}}
&& \hspace{-0.2cm}
\exp\left(
- i \int_0^{\pi} d\sigma\;  A^i(X^+)
 \, \partial_\sigma X^i \right)\;
e^{i\; p^+  (X^-_>- X^-_<)} \vert_{\sigma={\frac{\pi}{2}}} = \\
= & & \hspace{-0.1cm} \exp\left(
- i \int_0^{\pi} d\sigma\;  A^i\left( X^+(\sigma)
-{\frac{ip^+}{2}} {\rm log}
\Bigl(4\; {\rm sin}^2({\pi/ 2} -\sigma) \Bigr) \right)
 \, \partial_\sigma X^i \right)\;. \nonumber
\end{eqnarray}
[If the reader is worried  about our use of the above
identity, he or she can express the
 exponential of the integral as an
(infinite) product of infinitesimal exponentials,
 expand each one
of these in a Taylor series,  and then
sandwich  $1 = e^{-C}e^C$ in the appropriate
 places. The result
is indeed (\ref{rrs}).] Repeating once
 more the operation for the
second exponential in (\ref{scatter1}), and doing some trivial
algebra, we obtain~:
\begin{equation}\label{shif}
S_{0\to 0} = \frac{{\cal N}}{4}
\; \langle 0|
 \exp\left(
- i \int_0^{\pi} d\sigma\;
A^i \Bigl(X^+(\sigma) +{ip^+} \log \vert\tan \sigma\vert \Bigr)
 \, \partial_\sigma X^i \right)\;|\!| {\rm Dp}
 \rangle\!\rangle  \,.
\end{equation}
We can furthermore now
replace $X^+(\sigma)$  by its  zero mode $x^+_c$.
This is true because the non-zero modes of $X^+$ can move
freely to the left or right, and annihilate
the bra vacuum  either directly,  or after
`bouncing off' the state  $|\!| {\rm Dp} \rangle\!\rangle$\ .

\EPSFIGURE[ht]{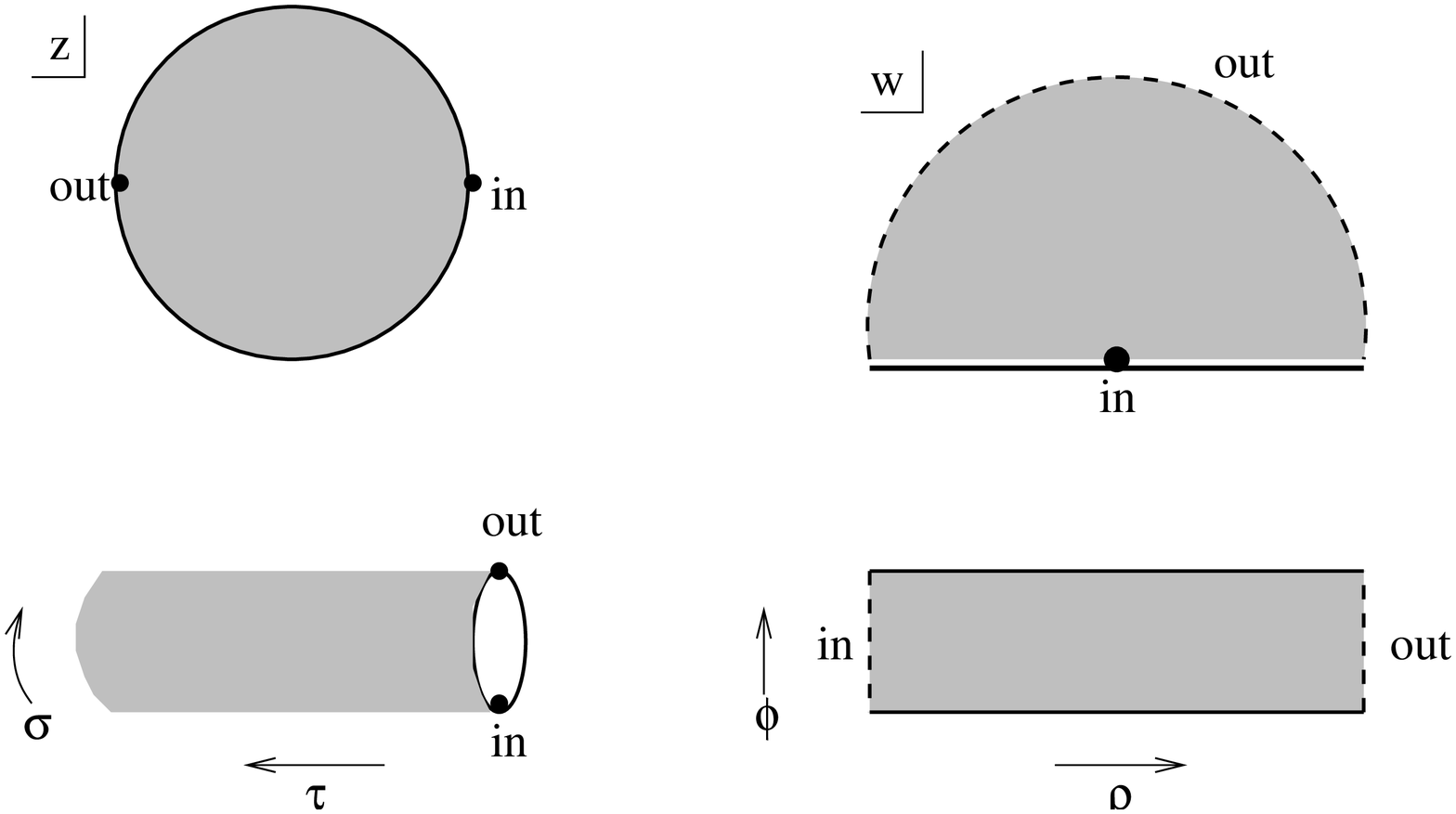,width=5.9in}
{The conformal transformation 
that maps the half-infinite cylinder
to the strip can be decomposed into three successive steps:
$\sigma+i\tau_E = -\frac{i}{2}\;{\rm log}z$, then  
$i w = (z-1)/(z+1)$ ,   and finally   
$\rho_E+i\phi = {\rm log}w$. 
Open-string vertex operators  are inserted
 at points `in' and `out'.
}

  Eq.~(\ref{shif}) should be compared to the expression
(\ref{ansatz}) for the boundary state. The main
difference is that the
argument of $A^i$ in the Wilson-loop operator has been  shifted
by the  imaginary amount ${ip^+} \log \vert\tan \sigma\vert$.
Since $A^i$  is a priori  defined
for real $x^+$, we need  to  evaluate the $\sigma$-integral 
by an appropriate  analytic continuation. Our  main point is
that {\it this analytic continuation is determined by the causal
structure of the Lorentzian open-string world-sheet.}
Indeed, what enters in  the argument of $A^i$
is the boundary restriction of the conformal transformation
\begin{equation}
\label{conftrans}
\rho_E + i\phi = {\rm log}({\rm tan}(\sigma+i\tau_E))
\end{equation}
that maps the Euclidean half-infinite cylinder,
  $\sigma\in [0,\pi)$ and
$\tau_E\in [0,\infty]$, 
to the Euclidean infinite strip,  $\phi\in [0,\pi]$ and $\rho_E\in
[-\infty, \infty]$; this  mapping 
is illustrated in figure 1. As $\sigma$ runs 
from $0$ to  $\pi$,  $\rho_E(\sigma)$ covers the  real axis
twice, corresponding to the two boundaries ($\phi=0$ and $\pi$)
of the strip.  The form of the
 argument in  (\ref{shif}) suggests
that the necessary analytic continuation should amount to a 
Wick rotation of  the  (preferred) open-string  time of the 
strip world-sheet.

To see how this works in detail, we
first apply the reasoning of section 3 to eliminate the
positive-frequency modes of $X^i$ and to arrive at an expression
like (\ref{fin}). Since  the negative-frequency  modes
annihilate the bra vacuum we find~:
\begin{eqnarray}\label{exxp}
S_{0\to 0} = \frac{\cal N}{4}\,
\langle 0\vert  \exp\Bigl(- \int_0^\pi d\sigma\;&&
\hskip -0.3cm  A^i\Bigl(x^+_c +ip^+\rho_E(\sigma)\Bigr)\;
 \int_0^\pi d\sigma^\prime \;
A^i\Bigl(x^+_c+ip^+ \rho_E(\sigma^\prime)\Bigr)\;\times
\nonumber\\
&&\times\;\partial_{\sigma}\partial_{\sigma^\prime}
D_E^{\rm cyl}(\sigma,\sigma^\prime)\Bigr)\;
|\!| {\rm Dp} \rangle\!\rangle\;.
\end{eqnarray}
We have here replaced the Feynman propagator on the cylinder
by its Euclidean counterpart,
 since the two coincide for space-like
separations. Furthermore, since the closed-string time
does not enter in (\ref{exxp})
 we may assume that the world-sheet
is Euclidean. Next we perform the conformal transformation
(\ref{conftrans}) which maps the
 integration contours to the oriented
boundary of the Euclidean strip. The transformation of the 
$\sigma$ and $\sigma^\prime$ derivatives
 in the above expression 
cancels precisely the two Jacobians  from the integration
measures. The  equal-time propagator 
on the cylinder is furthermore mapped
 to the boundary propagator 
of the Euclidean strip as follows~:
\begin{equation}\label{propsr}
D_E^{\rm cyl}(\sigma,\sigma^\prime)\  \rightarrow \ 
{1\over 2}D_E^{\rm str}(\rho_E-\rho_E^\prime,\phi_0) + f(\rho_E)
+ f(\rho_E^\prime)\  ,  
\end{equation}
where $f(x) = {1\over 4}\log\cosh x$,  and $\phi_0=0$ or $\pi$
for points on the same or opposite boundaries of the strip.
The relation (\ref{propsr}) 
is discussed in detail in the appendix. Plugging this 
in the expression (\ref{exxp}) gives~:
\begin{eqnarray}\label{exxpp}
S_{0\to 0} = \frac{\cal N}{4}\,
\langle 0\vert && \exp\Bigl(- \int_{-\infty}^\infty d\rho_E\;
  A^i\Bigl(x^+_c +ip^+\rho_E\Bigr)\;
 \int_{-\infty}^\infty d\rho_E^\prime \;
A^i\Bigl(x^+_c+ip^+ \rho_E^\prime\Bigr)\;\times
\nonumber\\
&&\times\;\partial_{\rho_E}\partial_{\rho_E^\prime}\left[
D_E^{\rm str}(\rho_E-\rho_E^\prime, 0)- 
D_E^{\rm str}(\rho_E-\rho_E^\prime, \pi)\right]   \Bigr)\;
|\!| {\rm Dp} \rangle\!\rangle\;.
\end{eqnarray}
Note that the extra
terms in the right-hand side of (\ref{propsr})
dropped  out upon derivation, while the relative
factor of $1/2$  
took   precisely care of the fact that as $\sigma$ runs over
$[0,\pi)$, the strip time covers the real axis twice.

  The expression (\ref{exxpp}) should
 be compared with  the result
of section 2~: $S_{0\to 0} = e^{i\delta_1-\delta_2}$ with 
$\delta_1+i\delta_2$ given by eq.~(\ref{pha}). 
The two results indeed agree if we 
 Wick rotate 
\begin{equation}
\rho_E\to -i\rho\ \  , 
\end{equation}
thereby sending the Euclidean to the Feynman propagator.  
The 
$x^+_c$ in the  arguments of the $A^i$ can be, 
of course, absorbed 
in the integration variables,  and we must also use our 
condition on  ${\cal N}$. 
This completes the proof  of open/closed-string duality for this
particular process. Other one-to-one
 S-matrix elements can be analysed
in a similar  way.

A subtle point in the above derivation   concerns possible 
singularities of the profile functions $A^i$ in the complex plane,
which could a priori obstruct the
 Wick rotation of the integration 
contour. 
Strictly speaking the right-hand-side of the BCH formula (\ref{bch1})
is defined by a Taylor expansion in powers of the commutator
$[C,B]$. As a result, 
the background fields in eq.~(\ref{exxp}) 
are also defined as Taylor series around the point on the real axis
$x^+_c$. The Wick rotation $\rho_E\to -i\rho$ 
makes each term in these expansions real. 
Singularities of $A^i$ in the complex plane
may  limit the radius of convergence of these series.
However these series still determine 
 unique continuous functions  on
the  integration axis which obviously coincide with the 
original wave profiles.\footnote{If the $A^i$  are
only piecewise continuous, we can  approximate them 
 by a sequence
of continuous functions.} 

In summary we note  that (as expected) 
 the Euclidean world-sheet
calculation gives the same result whether performed  in the open- 
or in the closed-string channel. The passage to a Lorentzian
world-sheet must, on the other hand, 
 be done by Wick rotating the
time coordinate of the open channel.
 This is the  choice that leads to a 
globally consistent causal structure on the string worldsheet. 
[The closed-channel time comes to an
 'abrupt' end on the disc boundary].
We turn  now  to the more general
 situation, in which there are
several incoming and/or outgoing strings.

\section{\Large Multi-string tree amplitudes}

   The boundary-state calculation of the previous section
can be generalized easily to  tree-level amplitudes
involving arbitrary numbers of open strings. 
 Let $i=1,\ldots, N$ label the incoming open strings while the
outgoing open strings are labelled by $i=N+1,\ldots,  N+M$.
 Each 
string carries a fraction $\alpha_i$
 of the conserved $p^+$ momentum~;
thus by construction we have 
$\sum_{i=1}^N \alpha_i = \sum_{j=N+1}^{N+M} \alpha_j=1$.
For simplicity we will take all these open strings
 to be in their
tachyonic ground state,  
and to have vanishing transverse momentum.

The amplitude (\ref{scattering}) is now replaced by a
more general expression, where incoming and
 outgoing open strings are
represented by vertex operators inserted at angular positions
$\sigma_i$ on the circle. Three of the $N+M$
 insertion points, 
for instance $\sigma_1,\sigma_2$ and $\sigma_3$,
 can be fixed using 
the  SL$(2,\RR)$ invariance of the disc. The 
total amplitude is  obtained by integrating
over the  remaining $N+M-3$ positions.
Since we consider here a single Dp-brane, we  do not  worry
about Chan-Paton matrices.

Following the same reasoning  as in 
section 4, one finds that this disc diagram equals
\begin{eqnarray}\label{exxpgen}
S_{{0\cdots 0} \rightarrow 0\cdots 0} = 
\left(\prod_{i=4}^{N+M} \int_0^\pi d\sigma_i\right) \; 
\prod_{i>j} && 
 e^{-p_i\cdot p_j D_E^{\rm cyl}(\sigma_i,\sigma_j) } \, 
\times 
\\
\times\langle 0\vert  \exp\Bigl(- \int_0^\pi d\sigma\;
 A^i\Bigl(x^+_c + i p^+\rho_E\Bigr)\;
&&\hskip -0.3cm \int_0^\pi d\sigma^\prime \;
A^i\Bigl(x^+_c + ip^+ \rho_E^\prime\Bigr)
\;\partial_{\sigma}\partial_{\sigma^\prime}
D_E^{\rm cyl}(\sigma,\sigma^\prime)\Bigr)\;
|\!| {\rm Dp} \rangle\!\rangle\;,\nonumber
\end{eqnarray}
where 
$\rho_E(\sigma)$ is now given by the following expression:
\begin{equation}\label{confm}
\rho_E(\sigma) =
\sum_{i=1}^N 2\alpha_i\, D_E^{\rm cyl} (\sigma,\sigma_i)
- \sum_{i=N+1}^{N+M} 2\alpha_i\, D_E^{\rm cyl} (\sigma,\sigma_i) \,,
\end{equation}
while $\rho_E^\prime$ is the same function with $\sigma$ replaced
by $\sigma^\prime$. 
The function defined by (\ref{confm}) is  the restriction to
the boundary of the map between the
 half-infinite cylinder with $N+M$
marked points on the boundary, and the open-string diagram in
light-cone gauge. The latter is an infinite
 strip with cuts which
extend either to $\rho_E= -\infty$ or to 
$\rho_E= +\infty$. This transformation is  illustrated for
$N=1$ and $M=2$ in figure 2.  

\EPSFIGURE[ht]{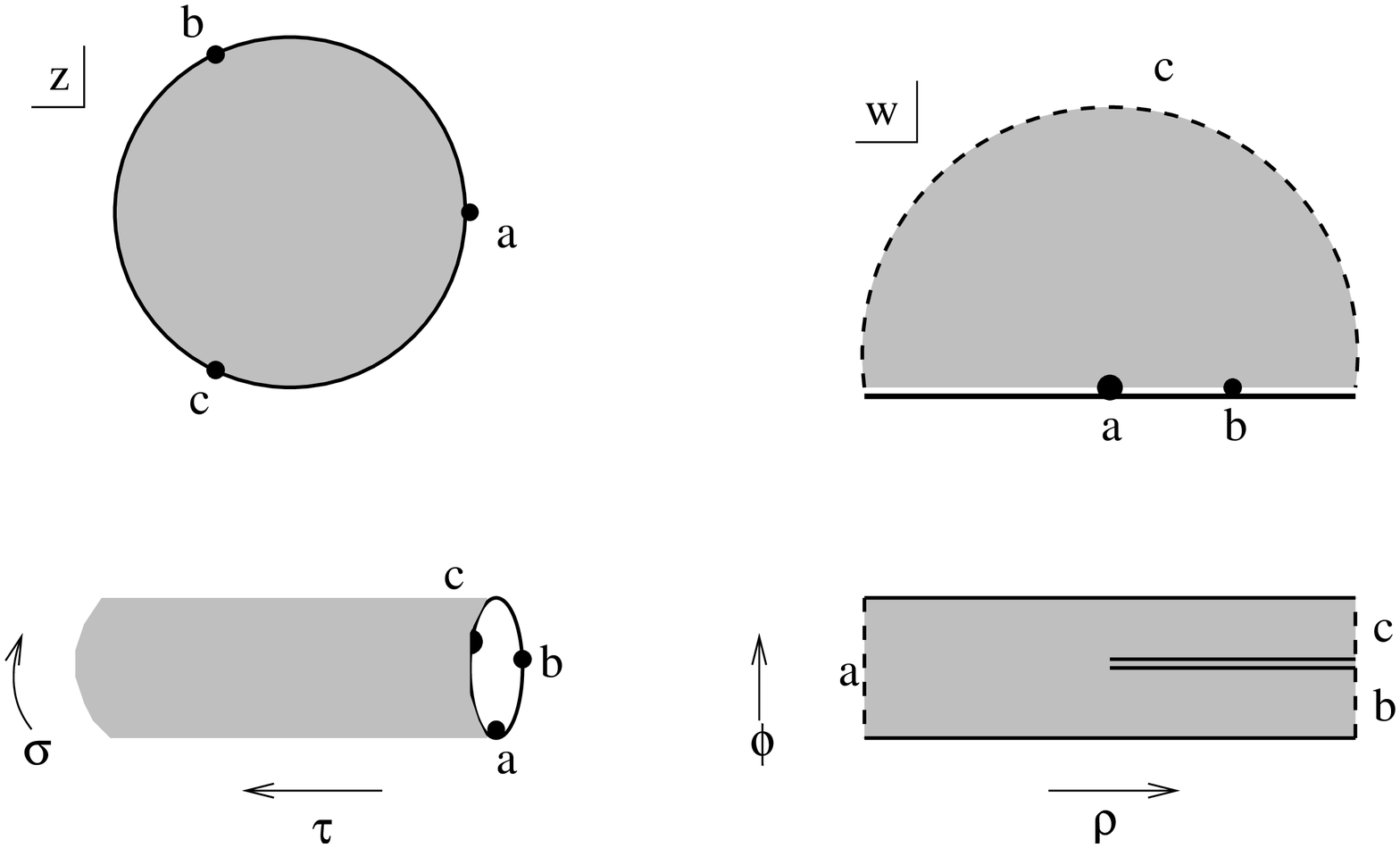,width=5.9in}
{The transformation (\ref{confm})
from the half-infinite cylinder to the strip
in the case of one incoming and two outgoing strings. The first 
two steps are the same as in figure 1, while the map from
the $w$ plane to the cut-strip   is
a `Schwarz-Christoffel transformation' (see for example
\cite{GSW}).}

  The rest of the calculation should
 then proceed as in the previous
section: one first changes integration
 variables from $\sigma$ to the
open world-sheet time $\rho_E$. All
  integration contours run
now around the oriented boundary of the cut-strip. 
Mass-shell conditions and momentum conservation should ensure that
Jacobians and extra terms in the transformed propagators cancel.
Finally, the form of the argument of
 the $A^i$ shows  that
the Wick rotation should be done in the time coordinate, $\rho_E$, 
of the cut strip.  We believe (but have not  checked explicitly) 
that for more general diagrams,
with loops and/or with asymptotic closed-string states,
it is always  the light-cone gauge worldsheet that singles  out 
the 'prefered',  Wick-rotated time.

  We postpone the explicit calculation
 of these amplitudes, including
the case where the emitted open strings are gauge bosons
or gravitons, to future work \cite{BCG}.
 A subtle point, to which we
hope to return,  concerns  the causal 
 propagator on a cut world-sheet. 
 The study of potential divergences of these amplitudes, and
more generally of the physics of back reaction, is especially
interesting for the reasons exposed  in ref. \cite{hull}.

\medskip

\section*{Acknowledgments}

We thank Nicolas Couchoud,
Michael Green, Nikita Nekrasov, Kostas Skenderis and
G\'erard Watts for useful conversations,
 and Jan Louis for hospitality
in the last stages of this work.  
MRG is grateful to
the Royal Society for a University Research Fellowship, and
acknowledges  partial support from the
 PPARC Special Programme Grant
PPA/G/S/1998/0061 `String Theory and Realistic Field Theory'.
CPB thanks the members of the theory group at  ETH for
their warm hospitality.
This research has been  partially supported by
  the European Networks
`Superstring Theory' (HPRN-CT-2000-00122) and
`The Quantum Structure of Spacetime'
(HPRN-CT-2000-00131).

\appendix

\section{\Large Relation between world-sheet propagators}

 In this appendix we establish explicitly the relation between
the two-point functions on the cylinder and on the  strip, 
which enters in the derivation of
 the single-open-string S-matrix from
the boundary state.

The Feynman propagator on the strip for
 a scalar field with Neumann
boundary conditions has the following standard representation~:
\begin{equation}\label{stripFF}
D_F^{\rm str}(\lambda,\phi)
= \sum_{n=-\infty}^{\infty}
\int \frac{d\omega}{2\pi}\; \frac{i}{\omega^2
- n^2 + i\epsilon}\;  e^{-i\omega\lambda +  i n\phi}\ .
\end{equation}
By performing the $\omega$ integration one recovers
 the expression
(\ref{stripF}) in the main text
 (for $\phi=0,\pi$). Wick rotating
the world-sheet time
 (\ie\ replacing $\omega\mapsto i\omega$ and
$\lambda\mapsto i\lambda_E$) leads to the Euclidean
propagator
\begin{equation}\label{stripFf}
D_E^{\rm str}(\lambda_E,\phi)
= \sum_{n=-\infty}^{\infty}
\int \frac{d\omega}{2\pi}\; \frac{1}{\omega^2
+ n^2 }\;  e^{i\omega\lambda_E +  i n\phi}\ .
\end{equation}
On the boundaries of the strip this reads (up to a constant)
\begin{equation}\label{stripff}
D_E^{\rm str}(\lambda_E,\phi_0)
= - \log \left\vert e^{\lambda_E/2}
 \mp e^{-\lambda_E/2} \right\vert  \ ,
\end{equation}
where the minus and plus signs correspond, respectively,
to  $\phi_0=0$ and $\phi_0=\pi$. 

  The Euclidean propagator (\ref{stripff}) should be compared
with the equal-time two-point function on the cylinder
\begin{equation}\label{cylFP}
D_E^{\rm cyl}(\sigma,\sigma^\prime) =
 -\frac{1}{4}\; {\rm  log}\;
\left( 4 \; {\rm sin}^2(\sigma-\sigma^\prime)\right) \ , 
\end{equation}
where $\lambda_E$ is 
to be identified with
 $\lambda_E=\rho_E(\sigma)-\rho_E(\sigma')$.
Note that for  equal times the Feynman 
and Euclidean propagators
coincide. 
Inverting the relation
$\rho_E +i\phi_0  = {\rm log(tan}\sigma)$,
 which maps the boundaries
of the half-infinite cylinder and the infinite strip,   gives
\begin{equation}
e^{-2i\sigma} = \frac{1\mp i e^{\rho_E}}{1\pm i e^{\rho_E}}\ ,
\end{equation}
where the choice of sign refers to the two
 possible values of $\phi_0$.
Plugging this into (\ref{cylFP}) and doing
 some straightforward
algebra one finds
\begin{equation}\label{last}
D_E^{\rm cyl} = -\frac{1}{2}
\log \left\vert e^{\lambda_E/2} \mp e^{-\lambda_E/2} \right\vert 
+ \frac{1}{4}{\log\cosh\rho_E}+\frac{1}{4}{\log\cosh\rho_E^\prime}\ . 
\end{equation}
This differs from (\ref{stripff}) in
 two ways: (i) there is a relative
factor of 2 between the open- and
 closed-string two-point functions
coming from the contribution of mirror images when the insertion
points approach the boundary; and (ii)
 there are extra terms, due to
the fact that $X$ is not a conformal
 primary field, and which will
drop out when one considers the primary field $\partial X$.

As is explained in the main text, 
the factor of $2$ is precisely what is needed to relate 
(\ref{exxp}) to  (\ref{exxpp}). Indeed, since 
$\rho_E(\sigma)=\rho_E(\pi-\sigma)$,
 the integrals in  (\ref{exxp})
cover the integration domain in  (\ref{exxpp})
 twice. Furthermore, since
only the derivatives of the propagators
 appear, the extra terms in
(\ref{last}) do not contribute. 
This proves, after a Wick rotation,  the desired equality
between the open- and the closed-channel calculations of the 
S-matrix.


\end{document}